\documentclass[%
 aip,
 amsmath,amssymb,
 reprint,%
]{revtex4-1}

\usepackage{graphicx}
\usepackage{dcolumn}
\usepackage{bm}

\usepackage[utf8]{inputenc}
\usepackage[T1]{fontenc}
\usepackage{mathptmx}
\usepackage{etoolbox}
\usepackage{xcolor} 
\definecolor{pink}{HTML}{FB3199} 
\usepackage[colorlinks,
            linkcolor=pink, 
            anchorcolor=pink, 
            citecolor=pink]{hyperref} 

\begin{document}

\preprint{AIP/123-QED}

\title{Spin-orbit interactions of the twisted random light}

\author{Benli Li}
\affiliation{School of Physics and Material Science, Nanchang University, Nanchang 330031, China;}
\affiliation{Jiangxi Provincial Key Laboratory of Photodetectors, School of Physics and Materials Science, Nanchang University, Nanchang 330031, China}

\author{Yahong Chen}
\email{yahongchen@suda.edu.cn}
\affiliation{School of Physical Science and Technology \& Collaborative Innovation Center of Suzhou Nano Science and Technology, Soochow University, Suzhou 215006, China}
\affiliation{Key Laboratory of Light Field Manipulation and System Integration Applications in Fujian Province, Minnan Normal University, Zhangzhou 363000, China}

\author{Weimin Deng}
\affiliation{School of Physics and Material Science, Nanchang University, Nanchang 330031, China;}
\affiliation{Jiangxi Provincial Key Laboratory of Photodetectors, School of Physics and Materials Science, Nanchang University, Nanchang 330031, China}

\author{Tongbiao Wang}
\affiliation{School of Physics and Material Science, Nanchang University, Nanchang 330031, China;}
\affiliation{Jiangxi Provincial Key Laboratory of Photodetectors, School of Physics and Materials Science, Nanchang University, Nanchang 330031, China}

\author{Lipeng Wan}
\email{Lwanoptics@ncu.edu.cn}
\affiliation{School of Physics and Material Science, Nanchang University, Nanchang 330031, China;}
\affiliation{Jiangxi Provincial Key Laboratory of Photodetectors, School of Physics and Materials Science, Nanchang University, Nanchang 330031, China}

\author{Tianbao Yu}
\affiliation{School of Physics and Material Science, Nanchang University, Nanchang 330031, China;}
\affiliation{Jiangxi Provincial Key Laboratory of Photodetectors, School of Physics and Materials Science, Nanchang University, Nanchang 330031, China}

\date{\today}

\begin{abstract}
The twist phase of random light represents a nontrivial two-point phase, endowing the field with orbital angular momentum. Although the mutual transition of the spin and orbit angular momenta of coherent light has been revealed, the relationship between spin-orbital angular momentum interaction (SOI) and the twist phase has remained unexplored. This is because of the stochastic nature of random light, making it challenging to explore the properties of angular momenta that rely on well-defined spatial and polarization structures. This study addresses this gap from the view of the asymmetry coherent-mode decomposition for twisted random light to gain insight into the intricate interplay between the twist phase and the SOI within a tight focusing system. Our findings reveal that spin and orbit angular momentum transitions occur in the tightly focused twisted random light beam, yielding the transverse spin density controlled by the twist phase. This effect becomes more pronounced when the spin of random light and the chirality of the twist phase are the same. Our work may find significant applications in optical sensing, metrology, and quantum optics.
\end{abstract}

\maketitle

Spin angular momentum (SAM) is inherently linked to circular polarization, representing the intrinsic angular momentum of light. In contrast, orbital angular momentum (OAM) is associated with the spatial structure of light's wavefront, characterized by helical phase fronts that carry quantized angular momentum proportional to the topological charge of the phase singularity \cite{barnett_optical_2001,PhysRevE.69.056613,doi:10.1080/09500349414550911,li2024prime}. The interplay between SAM and OAM has been a subject of considerable interest \cite{Bliokh:11,dorney_controlling_2019,PhysRevLett.96.163905}, as their coupling can give rise to unique light-matter interaction phenomena and complex beam dynamics, such as the spin-orbit interactions (SOI) of light. A notable advancement in understanding these properties was introduced in a high-numerical aperture (high-NA) system \cite{zhao_spin--orbital_2007,doi:10.1021/acsphotonics.1c01190}, where the transfer of SAM to OAM can indeed occur due to a geometric Berry phase. Although the coupling and physical relationship between SAM and OAM have been well described in fully coherent light fields, they remain elusive in the context of random optical fields \cite{PhysRevA.101.053825,PhysRevA.104.013516,wang2022effect,Wang:22,PhysRevA.106.063522,PhysRevA.109.043503}, especially those with twist phases. The complexity arises primarily due to the unconventional phase of two-point correlations within partially coherent beams \cite{Simon:930,Sundar:932}. Such correlations are significantly influenced by the structured coherence properties of the field, including those introduced by the rotationally invariant twist phase first proposed by Simon and Mukunda \cite{simon1993twisted}. This twist phase not only enriches the coherence structure but also encodes unique angular momentum characteristics, making it a cornerstone for understanding the SOI in random light. This innovative concept forms the basis for the twisted Gaussian Schell-model (TGSM) beam, offering a novel perspective on light fields with structured coherence and phase properties \cite{FOLEY1978297,FRIBERG1994127,Friberg1994InterpretationAE,PhysRevA.62.043816}. Research on the twist phase has attracted and advanced in recent years. These studies include the introduction of the controllable rotating Gaussian Schell-model beams \cite{Wan:19} and novel partially coherent vortex beams \cite{Ponomarenko:01,Mei:18,lipeng_wan_optical_2023}. However, the stochastic nature of random optical fields and the intricacies of the two-point correlation phase pose substantial challenges to fully elucidating the SOI of twisted random light, which hinders the exploration of the relationship between the twist phase and the SOI. 

Here we analyze the SOI of twisted random light beam in a high-NA system (i.e., during its tight focusing) by using the coherent-mode decomposition method \cite{gori_twisted_2015,Borghi:15}, revealing the interactions between the twist phase (OAM) and the polarization state (SAM). Such interactions significantly affect the spectral and spin density distribution of the tightly focused twisted random light. Remarkably, our study demonstrates that the SOI plays a pivotal role even in the context of reduced spatial coherence, offering transformative potential for applications in optical sensing, metrology, and quantum optics. By leveraging the SOI of random light, it becomes possible to achieve precise control and functionality in miniaturized optical devices, pushing the boundaries of subwavelength-scale technologies.

Let us start with the elementary twisted random light, i.e., a TGSM beam. The second-order statistical characteristics of the TGSM beam in the space-frequency domain are represented by the cross-spectral density (CSD) matrix \cite{simon1993twisted}
\begin{equation}
\mathbf{W}\left(\boldsymbol{\rho}, \boldsymbol{\rho}^{\prime}\right)=A \mathrm{e}^{-\frac{\boldsymbol{\rho}^2+\boldsymbol{\rho}^{\prime 2}}{4 \sigma^2}-\frac{\left(\boldsymbol{\rho}-\boldsymbol{\rho}^{\prime}\right)^2}{2 \delta^2}-{\mathrm{i} u\left(\boldsymbol{\rho}  \wedge \boldsymbol{\rho}^{\prime}\right)}} \hat{\mathbf{e}}^\ast(\boldsymbol{\rho})\hat{\mathbf{e}}^\mathrm{T}(\boldsymbol{\rho}^\prime),
\end{equation}
where $\boldsymbol{\rho} = (\rho_x, \rho_y)$ and $\boldsymbol{\rho}^{\prime} = (\rho_x^\prime, \rho_y^\prime)$ are two arbitrary points in the transverse plane of the beam, $A$ represents an inconsequential constant factor, which will be set to one henceforth, $\sigma$ and $\delta$ denote the beam width and the transverse coherence width, respectively, $\boldsymbol{\rho}  \wedge \boldsymbol{\rho}^{\prime} = \rho_x \rho_y^\prime - \rho_y \rho_x^\prime $, $u$ is the twist factor determining the strength of twist phase with its value bounded by $|u| \leq 1 / \delta^2$ to satisfy the positive semi-definiteness of the CSD function, $\hat{\mathbf{e}}$ denotes the unit polarization vector of the beam, whereas the asterisk and superscript T denote the complex conjugate and matrix transpose, respectively.

From the physical perspective of coherent-mode decomposition, the CSD matrix of a statistically stationary partially coherent source, regardless of its state of coherence, can be represented in the following form \cite{mandel1995optical}
\begin{equation}
\mathbf{W}\left(\boldsymbol{\rho}, \boldsymbol{\rho}^{\prime}\right)=\sum_{n,m} \lambda_{n,m} \boldsymbol{\Phi}_{n,m}^\ast(\boldsymbol{\rho}) \boldsymbol{\Phi}_{n,m}^{\mathrm{T}}\left(\boldsymbol{\rho}^{\prime}\right),
\label{eq2} 
\end{equation}
where $\lambda_{n,m}$ and $\boldsymbol{\Phi}_{n,m}(\boldsymbol{\rho})$ are the eigenvalues and eigenvectors for the CSD matrix, which can be obtained by solving the homogeneous Fredholm integral equation. We note $\lambda_{n,m}$ are non-negative due to the quasi-Hermitian definition of the CSD matrix. The pioneering works of Simon and Gori \cite{simon1993twisted,gori_twisted_2015} proved that the TGSM beam can be represented as incoherent superpositions of mutually orthogonal Laguerre--Gaussian (LG) modes with appropriate weights, i.e.,
\begin{equation}
\begin{aligned}
\boldsymbol{\Phi}_{n,m}(\boldsymbol{\rho}) = & \frac{1}{w} \sqrt{\frac{2 n !}{\pi(n+|m|) !}} \left(\frac{{\rho} \sqrt{2}}{w}\right)^{|m|} \\
 & \times L_n^{|m|}\left(\frac{2 {\rho}^2}{w^2}\right) e^{-{\rho}^2 / w^2} e^{i m {\varphi}} \hat{\mathbf{e}}(\boldsymbol{\rho}),
\end{aligned}
\end{equation}
\begin{equation}
\lambda_{n,m}=\frac{A \pi}{a+b+c / 2}\left(\frac{b+u / 2}{b-u / 2}\right)^{\frac{m}{2}}\left(\frac{a+b-c / 2}{a+b+c / 2}\right)^{\frac{|m|+2n}{2}}.
\label{eq5} 
\end{equation}
Above, $a=1 /\left(4 \sigma^2\right)$, $b=1 /\left(2 \delta^2\right)$, $c=2 (a^2+2 a b+u^2 / 4)^{1 / 2}$, $w=\sqrt{2/c}$, $n = 0, 1, \ldots$, $m  =0,\pm 1, \ldots$, $L_n^m$ is the Laguerre polynomials with radial and azimuthal indices being $n$ and $m$, respectively, and $(\rho, \varphi)$ are polar coordinates for $\boldsymbol{\rho}$. It can be seen from Eq.~\hyperref[eq5]{(4)} that the eigenvalues are closely related to the beam width $\sigma$, the transverse coherence width $\delta$, and the twist factor $u$.

\begin{figure}[t]
    \centering
    \includegraphics[width=1\linewidth]{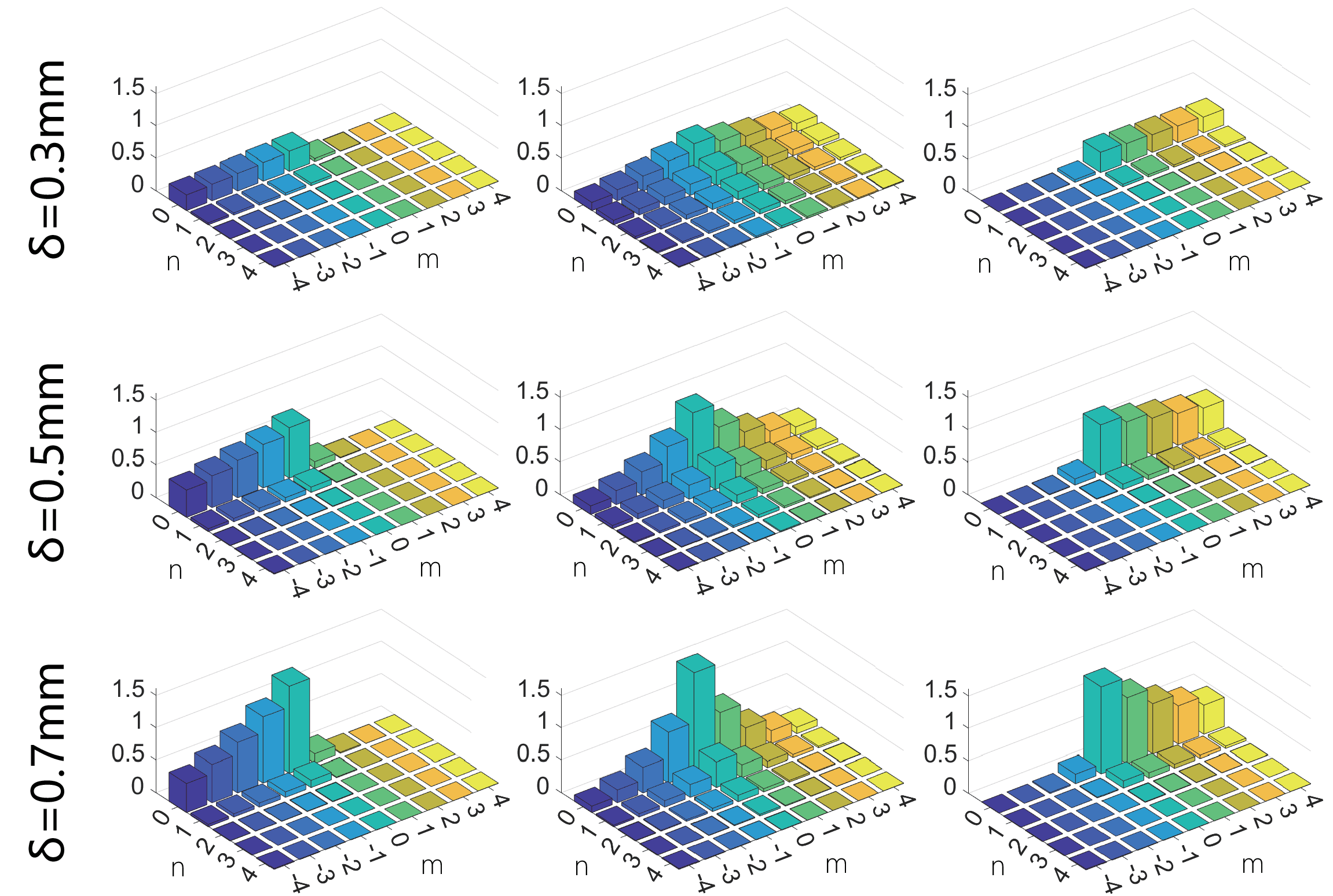}
    \caption{Distributions of the eigenvalues $\lambda_{n,m}$ along $n$ and $m$ for the TGSM beams with different $\delta$ and $u$. The first column $u\delta^2=-0.75$; the second column $u\delta^2=0$; and the third column $u\delta^2=0.75$.}
    \label{tzz}
\end{figure}

\hyperref[tzz]{Figure 1} illustrates the distribution of eigenvalues along $n$ and $m$ for different $\delta$ and $u$. As $\delta$ increases, the eigenvalue $\lambda_{0,0}$ of the fundamental mode increases significantly over that of other modes, reflecting the enhancement of the determinism of the beam. As shown in the second column of \hyperref[tzz]{Fig. 1}, when the twist factor $u=0$, the eigenvalues exhibit a symmetric distribution between $m=0$, i.e., $\lambda_{n,m}=\lambda_{n,-m}$. However, when the twist factor $u \neq 0$, we find the eigenvalues exhibit the asymmetric distributions, which will cause the nonzero OAM, and also the intensity rotation for the twisted beam during propagation \cite{Peng:18}. The sign of $u$ determines the direction for OAM and intensity rotation. It is also clear from \hyperref[tzz]{Fig. 1} that the eigenvalue decreases as the mode order $m$ or $n$ increases. Therefore, a finite number of coherent modes is sufficient for accurately representing the TGSM beam, in practical applications. In this work, we apply such coherent-mode decomposition to study the tight focusing properties, especially the SOI of the TGSM beam. The coherent modes with weight $\lambda_{n,m} / \lambda_{0, 0} \geq 0.01$ are used in our study for representing the TGSM beam. 

\begin{figure}[b]
    \centering
    \includegraphics[width=0.8\linewidth]{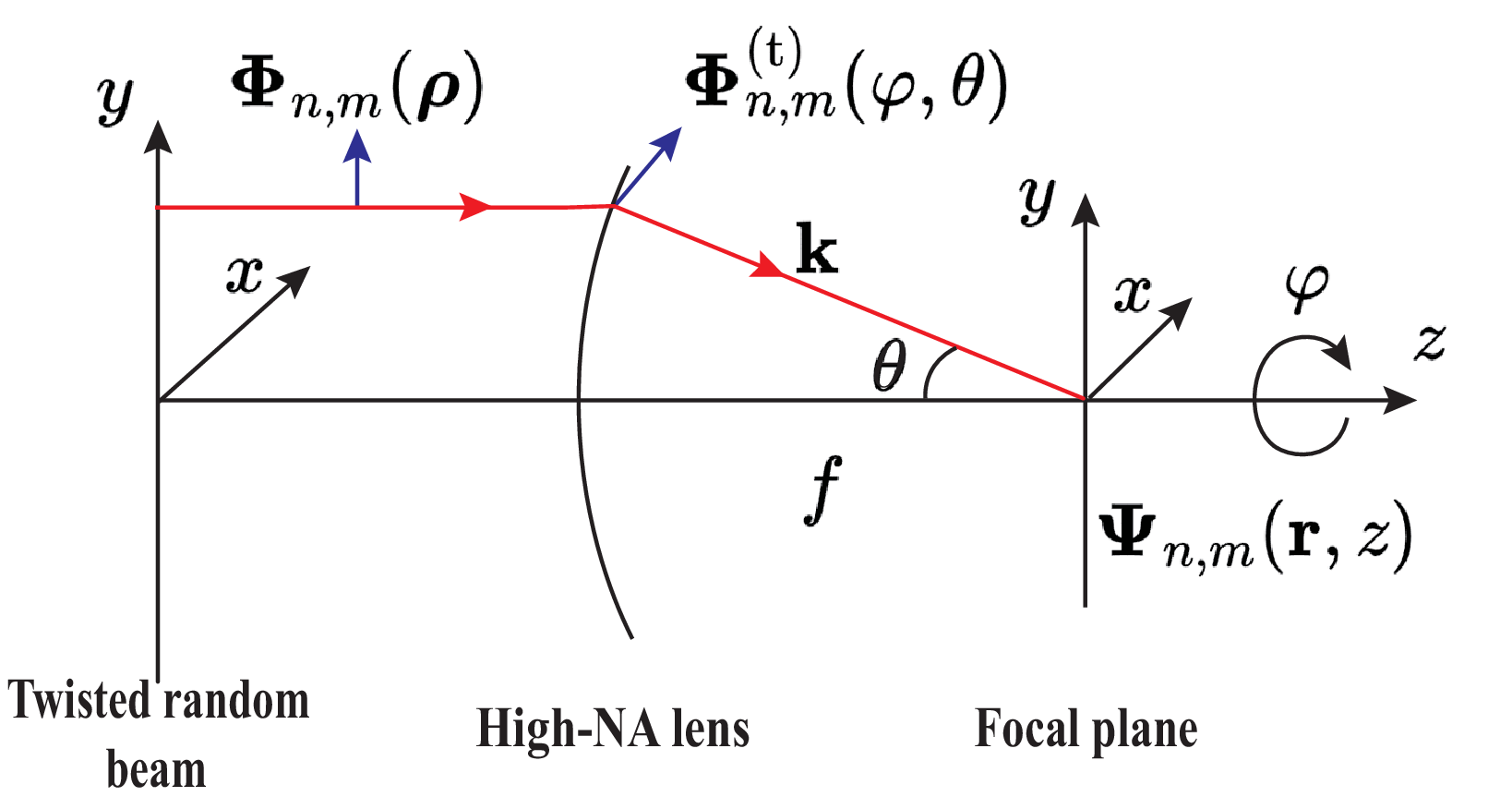}
    \caption{Geometry and notation related to tight focusing of a twisted random beam by a high numerical aperture objective lens with focal length $f$. In transmission through the objective the mode vector $\boldsymbol{\Phi}_{n,m}(\boldsymbol{\rho})$ transforms into $\boldsymbol{\Phi}_{n,m}^{(\text{t})}(\varphi,\theta)$ and the related wave vector $\mathbf{k}$ points towards the focal point and forms an angle $\theta$ with respect to the $z$ axis. } 
    \label{highNA}
\end{figure}

\hyperref[highNA]{Figure 2} shows the geometry of tight focusing of the TGSM beam through an aplanatic objective lens. We treat the transmission through the objective lens within the ray picture. The lens is assumed to obey the sine condition, i.e., the incident rays are refracted at the reference sphere of radius $f$ with $f$ being the focal length of the lens (see in \hyperref[highNA]{Fig. 2}). After the reference sphere, the input coherent mode $\boldsymbol{\Phi}_{n,m}(\boldsymbol{\rho})$ becomes $\boldsymbol{\Phi}_{n,m}^{(\text{t})}(\varphi,\theta)$ \cite{tong_fast_2020,leutenegger_fast_2006}. Here, the angles $\varphi$ and $\theta$ are shown in \hyperref[highNA]{Fig. 2}, and $(\varphi,\theta)$ represents a point on the reference sphere. We note a longitudinal ($z$) component appears in $\boldsymbol{\Phi}_{n,m}^{(\text{t})}(\varphi,\theta)$ due to that the radially polarized component of the input mode tilts at the off-axial points and acquires a longitudinal component. According to the Richards--Wolf method, the electric field near the focus is evaluated by the vectorial diffraction integral of the field over the spherical surface. Thus, the mode vector close to focus is obtained as \cite{tong_fast_2020}
\begin{equation}
\boldsymbol{\Psi}_{n,m}(\mathbf{r}, {z})=-\frac{\mathrm{i} f}{\lambda k} \mathcal{F}\left[\boldsymbol{\Phi}_{n,m}^{(\mathrm{t})}\left(k_x, k_y\right) \mathrm{e}^{\mathrm{i} k_z z} k_z^{-1}\right]_{k_x, k_y},
\label{eq8}
\end{equation}
where $\mathbf{r}=(x,y)$ is the transverse vector in the output plane, $z$ is the longitudinal distance from the focal point $(x,y,z)=(0,0,0)$, ${k_x, k_y,k_z}$ are the components of the wave vector $\mathbf{k}$ in the Cartesian coordinates along the $x, y,$ and $z$ directions, with $k=|\mathbf{k}|=2 \pi  / \lambda$, and $\mathcal{F}[\cdot]_{k_x, k_y}$ denotes the Fourier transform over the variables $k_x$ and $k_y$. We remark that $k_z=$ $\left(k^2-k_x^2-k_y^2\right)^{1 / 2}$, as a function of $k_x$ and $k_y$, must be taken into account in the Fourier transform. In Eq.~\hyperref[eq8]{(5)}, $\boldsymbol{\Phi}_{n,m}^{(\text{t})}(\varphi,\theta)$ is represented in terms of $k_x$ and $k_y$ due to the relations $k_x = -k \sin \theta \cos \varphi$ and $k_y = -k \sin \theta \sin \varphi$, as shown in the geometry in \hyperref[highNA]{Fig.~2}.

Adding up the mode vectors $\boldsymbol{\Psi}_{n,m}(\mathbf{r}, z)$ with Eq. \hyperref[eq2]{(2)} yields the CSD matrix of the tightly focused TGSM beam near the focal region:
\begin{equation}
\mathbf{W}( \mathbf{r}_1, z_1 ; \mathbf{r}_2, z_2 )=\sum_{n,m} \lambda_{n,m}\boldsymbol{\Psi}_{n,m}^{\ast}\left(\mathbf{r}_1, z_1\right) \boldsymbol{\Psi}_{n,m}^{\mathrm{T}}\left(\mathbf{r}_2, z_2\right).
\label{eq9}
\end{equation}
The spectral densities for three orthogonal components of the tightly focused TGSM beam are obtained as
\begin{equation}
 P_j(\mathbf{r},z) = W_{jj} ( \mathbf{r}, z ; \mathbf{r}, z ),
\end{equation}
where $j \in (x,y,z)$ and $W_{jj}$ are the diagonal components for the CSD matrix. In addition to examining the spectral density, we can further explore the SAM density to analyze the relation between the SOI and the twist phase. The (electric) SAM density vector for the random field can be obtained by \cite{PhysRevA.104.013516,PhysRevA.109.043503}
\begin{equation}
\mathbf{S}(\mathbf{r},z) \propto \operatorname{Im}\left \langle \mathbf{E}^\ast(\mathbf{r},z)  \times \mathbf{E}(\mathbf{r},z)  \right \rangle,
\label{eq10}
\end{equation}
where Im denotes the imaginary part and the angle brackets are the ensemble average over random field realizations $\mathbf{E}(\mathbf{r},z)$ for the tightly focused light. With the coherent-mode representation, the SAM density components can be obtained by
\begin{equation}
\begin{aligned}
& S_x \propto \operatorname{Im}\left[\sum_{n,m} \lambda_{n,m} (\psi_{n,m} )_y^*( \psi_{n,m} )_z\right], \\
& S_y \propto \operatorname{Im}\left[\sum_{n,m} \lambda_{n,m} (\psi_{n,m} )_z^*( \psi_{n,m} )_x\right],\\
& S_z \propto \operatorname{Im}\left[\sum_{n,m} \lambda_{n,m} (\psi_{n,m} )_x^*( \psi_{n,m} )_y\right],
\label{eq12}
\end{aligned}
\end{equation}
where $(\psi_{n,m} )_j$ with $j \in (x,y,z)$ represent the Cartesian components for $\boldsymbol{\Psi}_{n,m}(\mathbf{r},z)$.

\begin{figure}[t]
    \centering
    \includegraphics[width=1\linewidth]{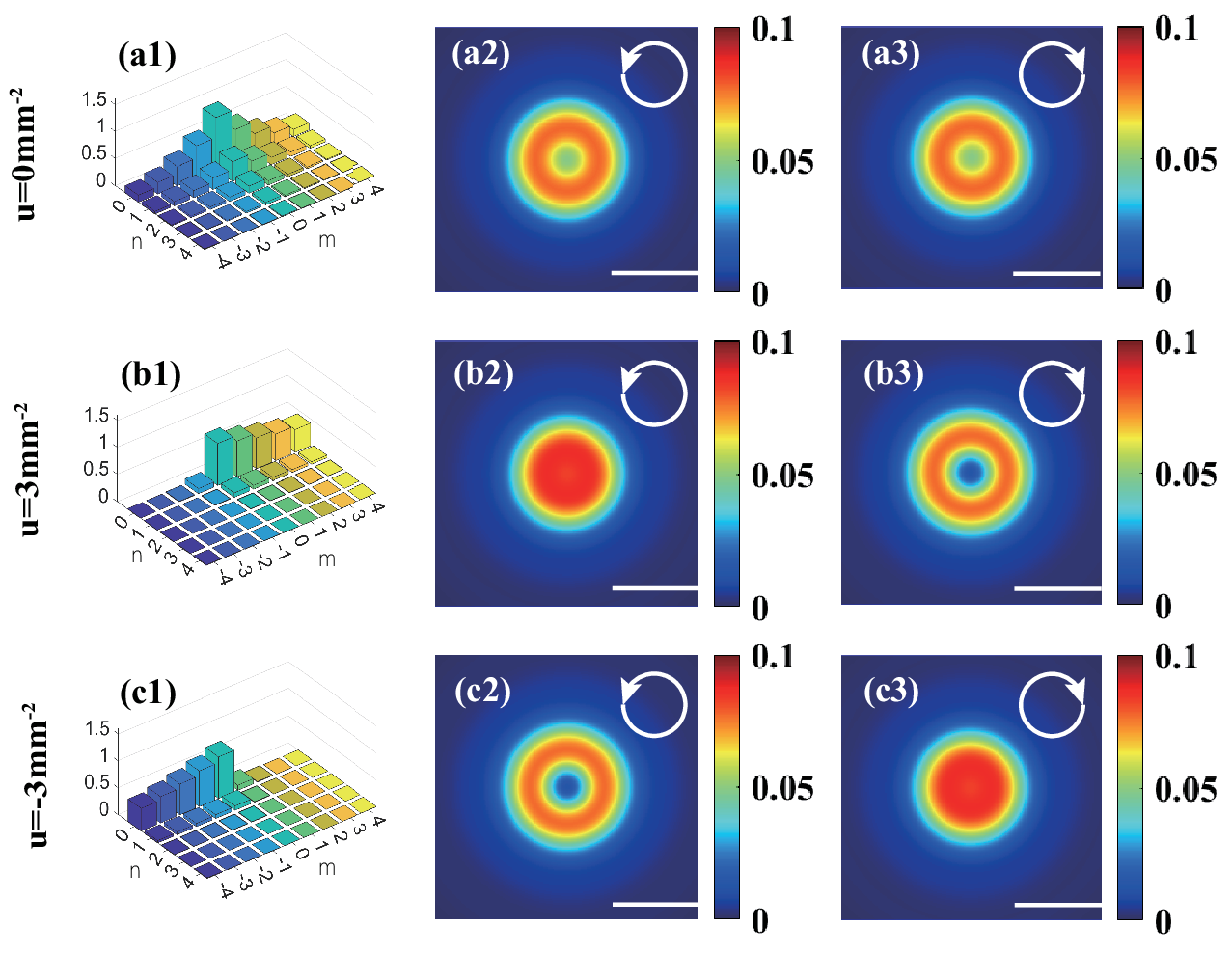}
    \caption{Eigenvalue distributions of the TGSM beams (first column), followed by the spatial distributions of spectral density $P_z$ for the longitudinal field component in the focal plane for the LCP (second column) and RCP (third column) TGSM beams in a high-NA system. The white circles with the counterclockwise arrow correspond to the LCP beams, whereas the circles with the clockwise arrow are associated with the RCP beams. (a1)--(a3), twist factor $u=0$, (b1)--(b3) twist factor $u=3 \ \text{mm}^{-2}$, and (c1)--(c3) twist factor $u=-3 \ \text{mm}^{-2}$.}
    \label{tu3}
\end{figure}

To explore the relationship between the twist phase and the SOI, we present simulation results of the spectral densities and the SAM density distributions in the focal plane. The parameters used in the simulation are: the wavelength $\lambda = 488$~nm, the beam width $\sigma=1$~mm, the coherence width $\delta=0.5$~mm, the numerical aperture of the lens $\mathrm{NA} = 0.9$, and the focal length is $f=0.5$~mm. All the plots are on a scale of $1.5~\mathrm{mm} \times 1.5~\mathrm{mm}$ with the scale bar representing one wavelength.

\begin{figure}[t]
    \centering
    \includegraphics[width=1\linewidth]{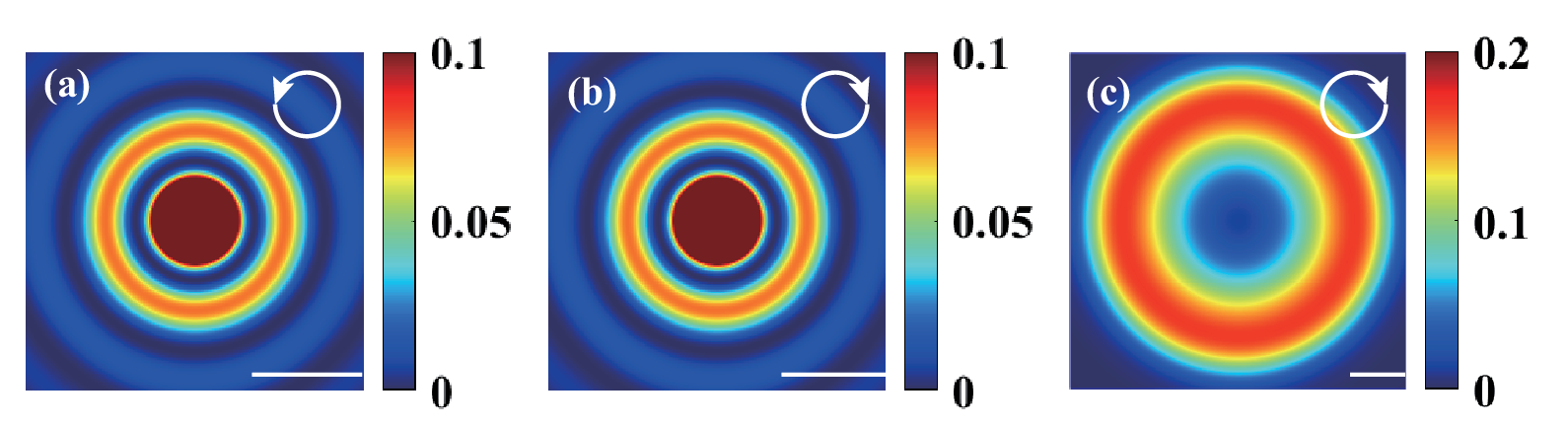}
    \caption{Spatial distributions of longitudinal spectral density $P_z$ in the focal plane for (a) the LCP $\text{LG}_{0,1}$ mode, (b) the RCP $\text{LG}_{0,-1}$ mode, and (c) the RCP TGSM beam of $u = 80 \ \text{mm}^{-2}$ and $\delta= 0.1 \ \text{mm}$ in a high-NA system.}
    \label{tu4}
\end{figure}

To investigate the spin-to-orbital conversion of the twisted random light beam under tight focusing, the incident beam is configured to be left circularly polarized (LCP) or right circularly polarized (RCP). \hyperref[tu3]{Figure 3} illustrates the spatial distributions of the spectral density ($P_z$) for the longitudinal component in the focal plane. In the case of non-twisted random beam (i.e., $u = 0 $), the spatial distribution of $P_z$ represents a quasi-donut pattern [see \hyperref[tu3]{Figs.~3(a2)} and \hyperref[tu3]{3(a3)}]. While the longitudinal component of the tightly focused twisted random beam ($u \neq 0$) appears as a prominent bright spot [see \hyperref[tu3]{Figs. 3(b2)} and \hyperref[tu3]{3(c3)}] or a central dark core [see \hyperref[tu3]{Figs. 3(b3)} and \hyperref[tu3]{3(c2)}], depending on the direction of SAM of the incident light. This change is attributed to the asymmetric mode distribution of the twisted random beam [see \hyperref[tu3]{Figs. 3(b1)} and \hyperref[tu3]{3(c1)}] and the interaction between the OAM and SAM carried by these modes. The twisted random beam with $u= 3 ~\text{mm}^{-2}$ consists mainly of positive vortex modes (i.e., the LG modes $\text{LG}_{0,0}$, $\text{LG}_{0,1}$, $\text{LG}_{0,2}$$,\dots$[see \hyperref[tu3]{Fig.~3(b1)}]). These modes carry $l\hbar$ SAM and $m\hbar$ OAM, where $l \in (-1,1)$ depending on the handedness of the circular polarization state and $m \in (0,1,2,\dots)$ depending on the topological charge of LG mode. Each mode of the random beam undergoes a transition from SAM to OAM in the high-NA system. The OAM for the longitudinal component of the tightly focused mode therefore transits into $p\hbar$ with $ p=m+l \in (-1, 0, 1,\dots)$ for $l=-1$ and $p \in (1,2,\dots)$ for $l=1$, which corresponds to the helical phase $\exp\left[ (m +l)i\varphi \right]$ for the longitudinal component \cite{zhao_spin--orbital_2007}. \hyperref[tu4]{Figure 4(a)} shows the $P_z$ of the LCP $\text{LG}_{0,1}$ mode ($l=-1$ and $m=1$), which does not exhibit the helical phase associated with the OAM ($p = 0$) due to cancellation of the OAM mode. In contrast, the other modes ($p\neq 0$) show a donut-like structure. Consequently, for twisted random beams, as the chirality of the twisted phase is opposite to that of the circular polarization, the focal plane retains a $p=0$ mode, resulting in the center of $P_z$ showing a bright spot. However, as the chirality of the twisted phase matches the handedness of the circular polarization, the helical phase associated with the OAM increases, leading to a dark-core pattern in $P_z$. This clear contrast allows us to observe the spin-to-orbital conversion. 

This analysis is further proved by considering a random beam of twist phase of opposite chirality $u = -3 \ \text{mm}^{-2}$. In this scenario, vortex modes $m<0$ contribute to the behavior of the beam [see \hyperref[tu3]{Fig. 3(c1)}]. For the RCP state, the $P_z$ component, as with the $\text{LG}_{0,-1}$ mode, demonstrates a bright spot [see \hyperref[tu3]{Figs. 3(c3)} and \hyperref[tu4]{4(b)}]. Conversely, for the LCP state, spin-to-orbital conversion can be observed [see \hyperref[tu3]{Fig. 3(c2)}]. Based on this interesting property, we can use it for spin-controlled transmission of light via chiral nano-aperture \cite{gorodetski2009observation} and AM-induced circular dichroism in non-chiral structures \cite{zambrana2014angular}. It is worth noting that the spin-to-orbital conversion is observable even though the degree of coherence is sufficiently low, provided that the twist factor is sufficiently large, as depicted in \hyperref[tu4]{Fig. 4(c)}. This is due to the increase in the beam's twist factor, which induces a pronounced asymmetric mode distribution in the composition of the beam.

Having investigated the spin-to-orbital conversion in twisted random light, we next investigate their SAM density and systematically explore how the variation of the OAM defined by the twist phase affects the longitudinal $z$ component of the SAM density ($ \textit{S}_\textit{z}$) in the focus of a high-NA system. 

\begin{figure}[t]
    \centering
    \includegraphics[width=1\linewidth]{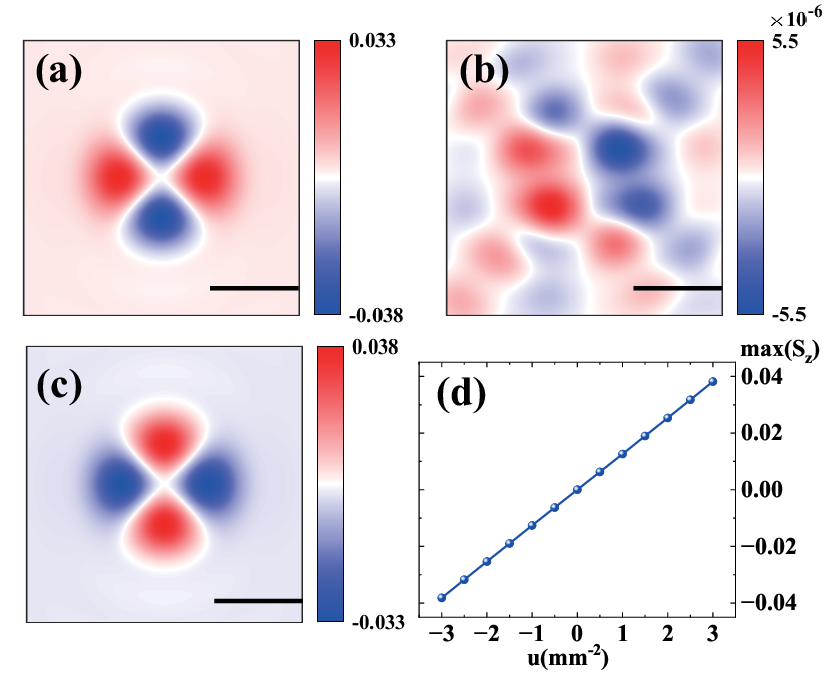}
    \caption{Focal-plane spatial distributions of the longitudinal SAM density $S_z$ for the linearly polarized TGSM beam with twist factor (a) $u=-3 \ \text{mm}^{-2}$, (b) $u=0$, and (c) $u=3  \ \text{mm}^{-2}$. (d) The maximum value of $S_z$ varies with $u$.}
    \label{linearly}
\end{figure}

\hyperref[linearly]{Figures 5(a)--(c)} illustrate the focal-plane spatial distribution of $ \textit{S}_\textit{z} $ for the linear polarized TGSM beam with $u=-3 \ \text{mm}^{-2}$, 0$ \ \text{mm}^{-2}$ and 3$ \ \text{mm}^{-2}$, respectively. The other parameters remain unchanged. Although the incident light carries zero SAM, we observe nonzero (both local and integral) SAM in the focus, in the case that the twist factor is nonzero. This highlights that the OAM of the twisted random light is partially converted into SAM. This process occurs because of the helical-phase-induced change in the polarization of the tightly focused field \cite{Yu:18}. The reversal of the twist factor $u$ means that the azimuthal indices (the sign of the vortex phase) of the modes that make up the TGSM beam are subsequently reversed. This change implies that the incident light carries exactly the opposite OAM thus leading to a change in the spin density distribution. By analyzing the magnitude of $S_z$ in \hyperref[linearly]{Figs.~5(a)} and \hyperref[linearly]{5(c)}, we observed that the total longitudinal SAM is not zero. Therefore, the twist phase of the incident light affects not only the local distribution of the SAM density but also the global SAM of the tightly focused beam \cite{wang2022effect}.

\begin{figure*}[t]
    \centering
    \includegraphics[width=1\linewidth]{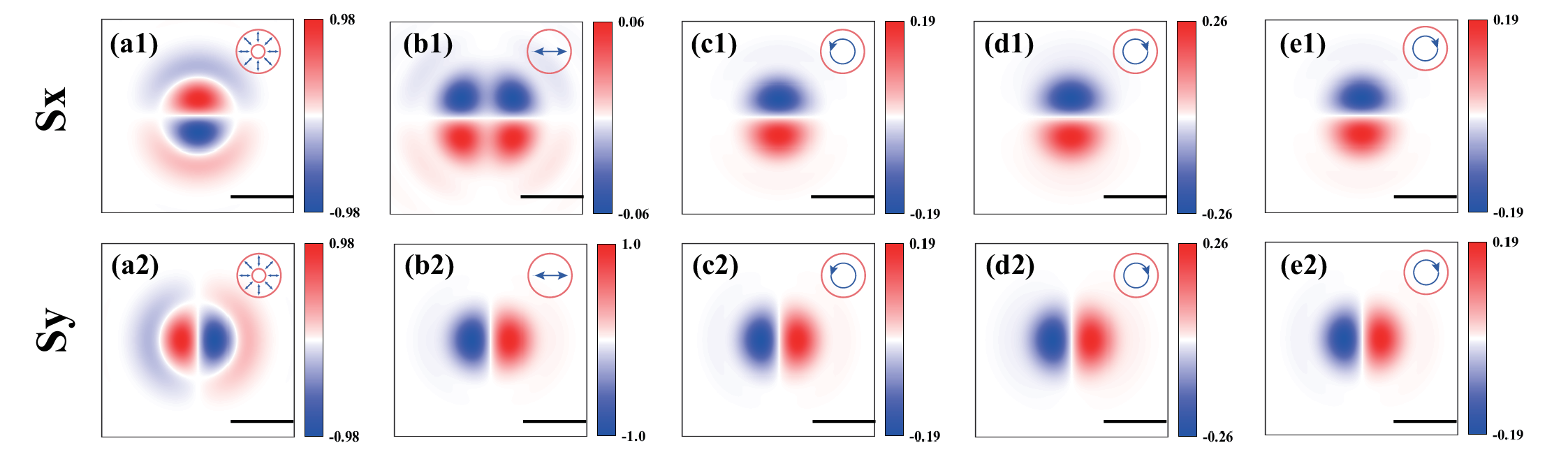}
    \caption{Focal-plane spatial distributions of the SAM density components $S_x$ and $S_y$ for the tightly focused TGSM beam with (a) a radial polarization state, (b) a linear $x$ polarization state, (c) an LCP state, and (d) and (e) an RCP state. In (a)--(d), the twist factor $u=3 \ \text{mm}^{-2}$, while in (e), the twist factor $u=-3 \ \text{mm}^{-2}$.}
    \label{tu6}
\end{figure*}

Notably, the longitudinal SAM density $\textit{S}_\textit{z}$ effectively vanishes in the absence of twist phase ($\mu=0$), as depicted in \hyperref[linearly]{Fig. 5(b)}. This behavior is consistent with the mode decomposition analysis: as the twist factor $u$ is zero, the eigenvalues of the modes are symmetrically distributed around the index $m = 0$ [see \hyperref[tu3]{Fig.~3(a1)}]. The fundamental mode $\text{LG}_{0,0}$ has zero OAM, does not contribute to $\textit{S}_\textit{z}$, while the contribution from higher-order modes with $m>0$ and with $m<0$ cancel each other out. \hyperref[linearly]{Figure 5(d)} further demonstrates the dependence of the maximum $\textit{S}_\textit{z}$ value on the twist factor $u$. As the value of the twist factor increases, the maximum longitudinal spin density component grows correspondingly. 

In addition to longitudinal spin, the high-NA system triggers the appearance of the longitudinal field component, resulting in transverse spin \cite{PhysRevLett.114.063901,Chen:17,bliokh_transverse_2015}. We study the transverse SAM density in the focal plane for the tightly focused twisted random beams with various initial polarization states, i.e., radial polarization state, linear polarization state, and left and right circular polarization states [see \hyperref[tu6]{Fig.~6}]. To ensure consistency of simulation conditions, the wavelength and beam width of the incident light, and the numerical aperture and focal length of the lens are kept constant.
 
In the case of the radial polarization state, the transverse SAM density in the focal plane exhibits the double-ring structure, which is distinct from the distributions for other polarization states [see \hyperref[tu6]{Figs.~6(a1)} and \hyperref[tu6]{6(a2)}], with the two rings displaying opposite spin directions \cite{wang2022effect}. For the linearly $x$-polarized incident beam, we find that $S_y$ is more pronounced than $S_x$ [see \hyperref[tu6]{Figs.~6(b1)} and \hyperref[tu6]{6(b2)}]. This phenomenon arises because of the emergence of the longitudinal field component which is out of phase with the transverse components. This phase difference enhances the SAM density in the specific directions \cite{neugebauer2018magnetic}.

When the input twisted random beam carries SAM (i.e., with circular polarization states), a similar spatial distribution of the transverse SAM density is observed in the focal plane for both handedness of circular polarization and opposite $u$. However, a subtle difference is noted in the magnitude of the SAM density [see the color bars in \hyperref[tu6]{Figs.~6(c1),~(c2),~(d1),~(d2),~(e1),~and~(e2)}]. For the LCP beam $(l=-1)$ whose handedness of circular polarization is opposite to the chirality of the twist phase $u=3 \ \text{mm}^{-2}$, the fundamental mode $\mathrm{LG}_{0,0}$ dominates due to spin-to-orbital conversion, as discussed above. The higher-order vortex modes with $|m|>0$ generate reversed transverse SAM components compared to the fundamental mode, resulting in the relatively low transverse SAM density [see \hyperref[tu6]{Figs.~6(c1) and (c2)}]. In contrast, for the RCP beam $(l=1)$ whose handedness of circular polarization coincides with the chirality of the twist phase $u=3 \ \text{mm}^{-2}$, the higher-order vortex mode dominates, resulting in the relatively high transverse spin [see \hyperref[tu6]{Figs. 6(d1) and (d2)}]. Thus, as the chirality of the twist phase coincides with the handedness of the circular polarization, the transverse SAM density is enhanced. This is further proved for the twist phase with $u = -3 \text{mm}^{-2}$, whose chirality is opposite to the handedness of the circular polarization [\hyperref[tu6]{Figs. 6(e1) and (e2)}], where the transverse spin decreases compared to \hyperref[tu6]{Figs. 6(d1) and (d2)}.

In a possible experimental setup, the required number of LG modes, for synthesizing the TGSM beam, is first determined based on pre-set parameters, such as the beam width and transverse coherence width. The corresponding holograms are then synthesized based on these modes. Next, the eigenvalues of each mode are treated as their occurrence probabilities, and the associated holograms are sequentially loaded onto a spatial light modulator (SLM) according to these probabilities. By time-averaging the light field output from the SLM, a TGSM beam \cite{zhang2021generating} with a controllable twist phase is ultimately obtained. To verify the SOI, the circularly polarized TGSM beam is tightly focused using a high-NA lens. To provide evidence of the SOI of the twisted random light beam during the tight focusing, we can observe the rotational velocity change of absorbing particles in the focal region \cite{bliokh_spinorbit_2015}. Alternatively, a nanoscale dipole-like particle can be employed as a field probe, positioned near the focal plane, to measure the local transverse and longitudinal SAM densities in the focal plane \cite{PhysRevLett.114.063901}.

In conclusion, this study provided a comprehensive analysis of the spin-orbit interaction (SOI) in twisted random light using the mode decomposition method. We have demonstrated that the twist phase plays a pivotal role in shaping the spectral and spin angular momentum (SAM) density distributions of the tightly focused twisted random beam. Our results reveal that the chirality of the twist phase, relative to the handedness of the circular polarization state, significantly influences the central spectral density features in the focal plane. Furthermore, we have shown that the focal-plane SAM density distribution is highly sensitive to the twist factor, with its maximum value increasing as the magnitude of the twist factor grows. These findings provide a fresh perspective on how the stochastic nature of partially coherent light can be harnessed to precisely control angular momentum properties, even in the absence of well-defined spatial or polarization structures. This work not only deepens our understanding of the underlying physics of SOI in random optical fields but also offers practical implications for applications such as optical manipulation and quantum information processing. By manipulating the twist phase, it becomes possible to finely tune the spin characteristics of light, enabling more precise beam control and efficient information transfer. Our findings pave the way for future research into the intricate interplay between spin, orbital angular momentum, and coherence in random light fields, opening up new possibilities for developing advanced optical technologies.

\begin{acknowledgments} 
This research was supported by National Natural Science Foundation of China (Grant Nos. 62305146, 12164027, 12274310); Training Program for Academic and Technical Leaders of Major Disciplines in Jiangxi Province (Grant Nos. 20243BCE51163, 20243BCE51145); Natural Science Foundation of Jiangxi Province (Grant Nos. 20232BAB211031, 20242BAB20023); Nanchang University Youth Training Program (Grant No. PYQN20230064); Project of Preeminent Youth Fund of Jiangxi Province (Grant No. 20224ACB211002); Jiangxi Provincial Key Laboratory of Photodetectors (Grant No. 2024SSY03041); Key Laboratory of Light Field Manipulation and System Integration Applications in Fujian Province (Grant No.~GCTK202305).
\end{acknowledgments}

\bibliography{aipsamp}

\end{document}